\documentclass[twocolumn,prb,aps,showpacs]{revtex4}
\usepackage{graphics}
\usepackage{dcolumn}
\usepackage{bm}
\usepackage{amsmath}
\usepackage{amssymb}
\usepackage{epsfig}
\usepackage{pictex}
\begin{document}
\title{ Spin waves in the block checkerboard antiferromagnetic phase}
\author{Feng Lu$^{1}$, and Xi Dai$^{1}$}
\affiliation{\it
 $^1$  Beijing National Laboratory for Condensed Matter Physics,
and Institute of Physics, Chinese Academy of Sciences, Beijing 100190, China
}
\date{today}

\begin{abstract}
Motivated by the discovery of new family 122 iron-based superconductors,
we present the theoretical results on the ground state phase
diagram, spin wave and dynamic structure factor of the extended $J_{1}-J_{2}$
Heisenberg model. In the reasonable physical parameter region of $K_{2}Fe_{4}Se_{5}$
, we fi{}nd the block checkerboard antiferromagnetic order phase is
stable. There are two acoustic branches and six optical branches spin
wave in the block checkerboard antiferromagnetic phase, which has
analytic expression in the high symmetry points. To compare the further
neutron scattering experiments, we discuss the saddlepoint structure
in the magnetic excitation spectrum and calculate the predicted inelastic
neutron scattering pattern based on linear spin wave theory.
\end{abstract}


\maketitle

\section{Introduction}
%

Searching high-$T_c$ superconductivity has been one of the central topics
in condensed matter physics\cite{Lee-P-A-1}. Following the discovery
of the copper-based superconductors two decades ago\cite{Bednorz-1},
the second class of high-transition temperature superconductors has
been reported in iron-based materials\cite{Kamihara-1,Chen-G-F-1,Lei-H-1}.
The iron pnictides contains four typical crystal structures, such as 1111-type
ReFeAsO (Re represents rare earth elements)\cite{Kondo-1}, 122 type
Ba(Ca)Fe$_{2}$As$_{2}$\cite{Wray-1,Xu-Gang-1}, 111 type LiFeAs\cite{Borisenko-1}
and 11 type FeSe\cite{Khasanov-1}.
The parent compounds with all structures except the 11 type
have the stripe like antiferromagnetic state as the ground state.
By substituting a few percent of O with F \cite{Kondo-1}or Ba with
K\cite{Wray-1,Xu-Gang-1}, the compounds enter the superconducting
(SC) phase from the SDW phase below $T_c$ .
In addition to the above four crystal structures, a new family of iron-based
superconducting materials with 122 type crystal structure have recently
been discovered with the transition temperature $T_c$ as high
as 33 K \cite{Guo-2}. However, these new "122" compounds differ from
the iron superconductors with the other structures in many aspects\cite{Maziopa-1}.
Firstly, the parent compound of the new "122" material is proposed to be $K_{0.8}Fe_{1.6}Se_{2}$
with intrinsic $root 5 by root 5 Fe$ vacancy ordering determined by various
experiments\cite{Wei-Bao-1,Wei-Bao-2,Wei-Bao-3}.
Secondly, the ground state for $K_{0.8}Fe_{1.6}Se_{2}$ is Mott insulator with
block antiferromagnetic order, which is observed by the neutron diffraction experiments\cite{Wei-Bao-1,Wei-Bao-2,Wei-Bao-3}.
By the first principles calculation, Yan et.
al\cite{Y-Z-Lu-1} and Cao et. al\cite{Dai-J-H} also found the block antiferromagnetic order is the most
stable ground state for $K_{0.8}Fe_{1.6}As_{2}$.

To date, a number of researches have been carried out to study the nature
of superconductivity and magnetic properties of these materials. The neutron scattering
experiments by Bao et.al\cite{Wei-Bao-3} have found the ground state of  $K_{0.8}Fe_{1.6}As_{2}$
to be block anti-ferromagnetic with magnetic moment around $3.4\mu_B$.
It has been proposed by several authors that the magnetic and superconducting
instabilities are strongly coupled together and the properties of magnetic excitations, such as spin wave,
play very crucial roles for the superconductivity in this family materials.
Zhang et. al\cite{Zhang-G-M-1} even suggested that the superconducting pairing
may be  mediated by coherent spin wave excitations in these materials.

In order to give the qualitative insight into the magnetic excitation properties
in this system, we studied the spin wave spectrum using
the Heisenberg model on the 2D square lattice with $\sqrt{5}\times\sqrt{5}$
vacancy pattern. There are four independent parameters are used in the model,
which correspond to the nearest neighbor and next nearest neighbor coupling between
spins.
We first obtain the ground state phase diagram as the function
of those parameters, based on which we calculate the spin wave spectrum as well as
the spin dynamic structure factor using the Holstein-Primakov transformation.
Our results demonstrate
that the block checkerboard antiferromagnetic order is stable in a wide range of
phase regime and there are two acoustic branches as well as
six optical branches spin wave in this system, which can be measured by the future neutron
scattering experiments.

\section{Model and Method}
\label{secmodel}
\begin{figure}[tp]
\vglue -0.6cm \scalebox{1.150}[1.15]
{\epsfig{figure=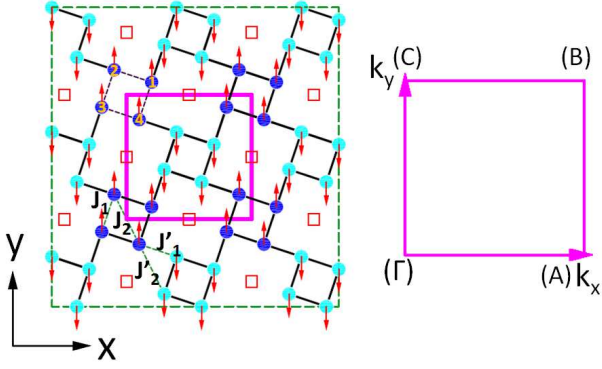,width=5.cm,angle=90}} \caption{
(color online)Schematic diagram of the 2 dimensional crystal and magnetic
structure (single layer for $K_{2}Fe_{4}Se_{5}$ ) and the corresponding
Brillouin Zone. On the left is the crystal structure and spin pattern
of the pnictides $K_{2}Fe_{4}Se_{5}$ . The up-arrow(blue) and down-arrow(azure)
atoms indicate the Fe-atoms with positive/negative magnetic moment,
respectively. Here we show the considered model with nearest neighbor
coupling $J_{1},J_{1}'$ and next nearest neighbor coupling $J_{2},J_{2}'$.
The coupling within each block is $J(J_{1},J_{2})$, and the coupling
between blocks is $J'(J_{1}',J_{2}')$. On the right is the positive
quadrant of the square-lattice Brillouin Zone showing wave-vectors
$\Gamma=(0,0)$, $A=(\frac{2\pi}{\sqrt{10}a},0)$,
$B=(\frac{2\pi}{\sqrt{10}a},\frac{2\pi}{\sqrt{10}a})$,
$C=(0,\frac{2\pi}{\sqrt{10}a})$. With the high symmetry line, the
spin wave are contained along the direction $\Gamma-A-B-\Gamma-C-B$.
The purple solid line marks the magnetic unit cell. The iron vacancy
site Fe is marked by the open square, and the occupied site Fe is
marked by solid circle with the blue or reseda color indicating spin
up or spin down.}
\label{fig:lattice}
\end{figure}

The simplest model that captures the essential physics in Fe-vacancies
ordered material $K_{2}Fe_{4}Se_{5}$ can be described by the extended
$J_{1}-J_{2}$ Heisenberg model on a quasi-two-dimensional lattice
\cite{Dai-J-H,Y-Z-Lu-1},
\begin{eqnarray}
H & = & J_{1}\sum_{i,\delta,\delta' (> \delta)}\overrightarrow{S}_{i,\delta}\cdot
                                     \overrightarrow{S}_{i,\delta'}
     \nonumber\\
 && +J'_{1}\sum_{i,\gamma,\delta,\delta'}\overrightarrow{S}_{i,\delta}\cdot
                                      \overrightarrow{S}_{i+\gamma,\delta'}
     \nonumber\\
 && +J_{2}\sum_{i,\delta,\delta'' (> \delta)}\overrightarrow{S}_{i,\delta}\cdot
                                      \overrightarrow{S}_{i,\delta''}
     \nonumber\\
 && +J'_{2}\sum_{i,\gamma,\delta,\delta''}\overrightarrow{S}_{i,\delta}\cdot
                                      \overrightarrow{S}_{i+\gamma,\delta''}
\end{eqnarray}
Here,  $\delta=1,2,3,4$ and $\gamma=1,2,3,4$;
the first and second terms represent nearest-neighbor (n.n.)
spin interactions in the intra- and inter- block, respectively, as
shown in Fig.\ref{fig:lattice}. The third and forth term are second-neighbor
( n.n.n.) spin interactions which are taken to be independent on the
direction in the intra- and inter- block. Here, i is the block index,
$\gamma$ denotes the nearest-neighbor block of i block. $\delta'$($\delta''$)
represents the site-index which is n.n (n.n.n) site of site $\delta$.
$J_{1}$ and $J'_{1}$ ($J_{2}$ and $J'_{2}$ ) indicate n.n. ( n.n.n.)
couplings of intra- and inter-block, respectively, which are illustrated
in Fig. 1. Here, we define $J_{2}$ is the energy unit.

In order to understand this $J_{1}-J_{2}$ Heisengerg Hamiltonian,
we depict the typical block spin ground state and the 2D Brillouin
zone (BZ) in Fig.\ref{fig:lattice}. The q-vectors used for defining
the high symmetry line of the spin wave is also depicted. Each magnetic
unit cell contains eight Fe atoms and two Fe vacancy which are marked
by the open squares, as shown in Fig.\ref{fig:lattice}. The block
checkerboard antiferromagnetic order are recently observed by the
neutron diff{}raction experiment in the $K_{2}Fe_{4}Se_{5}$ material.
An convenient way to understand the antiferromagnetic structure of
$K_{2}Fe_{4}Se_{5}$ is to consider the four parallel magnetic moments
in one block as a supermoment; and the supermoments then form a simple
chess-board nearest-neighbor antiferromagnetic order on a square lattice,
as seen from Fig.\ref{fig:lattice}a.
The distance between the nearest-neighbor iron is defined as $a$.
Then, the crystal lattice constant for the magnetic unit cell
is $\sqrt{10}a$. Fig.\ref{fig:lattice}b is the 2D BZ for the magnetic
unit cell.

We use Holstein-Primakoff (HP) bosons to investigate the spin wave
of the block checkerboard antiferromagnetic ground state. As we know,
linearized spin wave theory is a standard procedure to calculate the
spin wave excitation spectrum and the zero-temperature dynamical structure
factor\cite{Auerbach-1,D-X-Yao-1}. Firstly, we use HP bosons to replace
the spin operators, as shown in Appendix A.

Using Holstein-Primakoff transformations, we can obtain the HP boson
Hamiltonian,
\begin{eqnarray}
H & = & \sum_{k}\psi_{k}^{\dagger}\mathcal{H}_{k}\psi_{k}+E_{0}-N\cdot E_{\left(k=0\right)}
\end{eqnarray}
Here, $\begin{smallmatrix} \psi_{k}^{\dagger} & =
\end{smallmatrix}$
$\bigl( \begin{smallmatrix}
a_{k1}^{\dagger} & a_{k2}^{\dagger} & a_{k3}^{\dagger} & a_{k4}^{\dagger} & b_{-k-1}
& b_{-k-2} & b_{-k-3} & b_{-k-4}
\end{smallmatrix}\bigr)$;
$E_{0}$ is the classical ground state energy for block checkerboard
antiferromagnetic order, $E_{0}=8J_{1}NS^{2}-4J'_{1}NS^{2}+4J_{2}NS^{2}-8J'_{2}NS^{2}$.
The Specific expression for $\mathcal{H}_{k}$ is shown in Appendix B.

In the real space, we define the 'molecular orbital' , which is the
combination of HP boson operators in one block.
\begin{eqnarray}
\begin{smallmatrix}
\alpha_{i1} & = & \frac{1}{\sqrt{4}}\left(a_{i1}-a_{i2}+a_{i3}-a_{i4}\right)
    \nonumber\\
\alpha_{i2} & = & \frac{1}{\sqrt{4}}\left(a_{i1}+a_{i2}-a_{i3}-a_{i4}\right)
    \nonumber\\
\alpha_{i3} & = & \frac{1}{\sqrt{4}}\left(a_{i1}-a_{i2}-a_{i3}+a_{i4}\right)
    \nonumber\\
\alpha_{i4} & = & \frac{1}{\sqrt{4}}\left(a_{i1}+a_{i2}+a_{i3}+a_{i4}\right)
    \nonumber\\
\beta_{i-1} & = & \frac{1}{\sqrt{4}}\left(b_{i-1}-b_{i-2}+b_{i-3}-b_{i-4}\right)
    \nonumber\\
\beta_{i-2} & = & \frac{1}{\sqrt{4}}\left(b_{i-1}+b_{i-2}-b_{i-3}-b_{i-4}\right)
    \nonumber\\
\beta_{i-3} & = & \frac{1}{\sqrt{4}}\left(b_{i-1}-b_{i-2}-b_{i-3}+b_{i-4}\right)
    \nonumber\\
\beta_{i-4} & = & \frac{1}{\sqrt{4}}\left(b_{i-1}+b_{i-2}+b_{i-3}+b_{i-4}\right)
\end{smallmatrix}
\end{eqnarray}
Here, the $-$ represents the spin down block. The corresponding physical picture for each 'molecular orbital' is
shown in the Fig. 2.

\begin{figure}[tp]
\vglue -0.6cm \scalebox{1.150}[1.15]
{\epsfig{figure=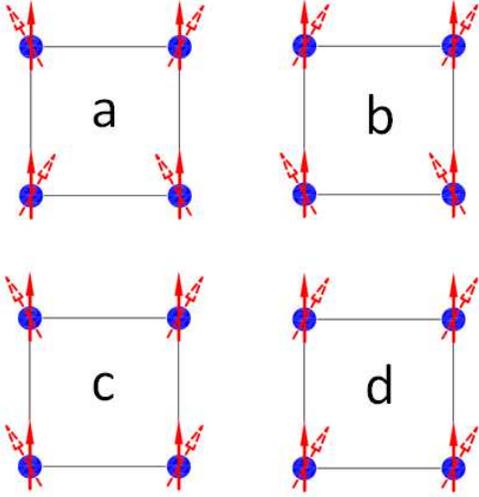,width=6.cm,angle=0}} \caption{
(color online) The schematic diagram for the 'molecular orbital' in the $\Gamma$ point:
 (a) The deviation of spin in site 1 and 3
has the same phase for the corresponding wave function; and the deviation
of spin in site 2 and 4 has the same phase for the corresponding wave
function. But the different between the phase of wave function for
site 1 and 2 is 180 degrees. (b)The deviation of spin in site 1 and
2 has the same phase for the corresponding the wave function; and
the deviation of spin in site 3 and 4 has the same phase for the corresponding
wave function. But the different between the phase of wave function
for site 1 and 3 is 180 degrees. (c) The deviation of spin in site
1 and 4 has the same phase for the corresponding wave function; and
the deviation of spin in site 2 and 3 has the same phase for the corresponding
wave function. But the different between the phase of wave function
for site 1 and 2 is 180 degrees. (d) The deviation of spin in site
1, 2, 3 and 4 all has the same phase for the corresponding wave function.}
\label{fig:mocular orbital}
\end{figure}

In the 'molecular orbital' basis, the Hamiltonian becomes,
\begin{eqnarray}
H_{k} & = & \sum_{k}\psi_{k}^{o \dagger}\mathcal{H}^{orbital}_{k}\psi_{k}^{o}
\end{eqnarray}
Here, $\begin{smallmatrix} \psi_{k}^{o \dagger} & =
\end{smallmatrix}$
$\bigl( \begin{smallmatrix}
\alpha_{k1}^{\dagger} & \alpha_{k2}^{\dagger} & \alpha_{k3}^{\dagger}
& \alpha_{k4}^{\dagger} & \beta_{-k-1} & \beta_{-k-2} & \beta_{-k-3} & \beta_{-k-4}
\end{smallmatrix}\bigr)$.
The matrix elements for different 'molecular orbital' in the same
block is zero, which is interesting; and we discuss the Hamiltonian
later. The Specific expression for $\mathcal{H}^{orbital}_{k}$ is shown in
Appendix B.

Because the boson Hamiltonian are big, one must use numerical diagonalization
to solve eigenvalues for spin wave, which has a standard procedure\cite{M-W-Xiao,M-W-Xiao-1,M-W-Xiao-3,M-W-Xiao-4,M-W-Xiao-5},
as shown in Appendix C.

To further understand the physical properties in this spin system,
we obtain the analytical eigenvalues and eigenvectors in some special
k points, such as $k=(0,0)$ and $k=(\frac{\pi}{\sqrt{10}a},\frac{\pi}{\sqrt{10}a})$ point.
In the $\Gamma$
point, the 'molecular orbital' Hamiltonian becomes four $2\times2$
block matrix
$\bigl( \begin{smallmatrix}
h_{m} & h_{m,m+4}\\
h_{m,m+4} & h_{m}\end{smallmatrix}\bigr)$
about $\alpha_{m}$ and $\beta_{-k-m}^{\dagger}$ ($m=1,2,3,4$), which
indicates that the 'molecular orbital' are intrinsic vibration modes
for $\Gamma$ point. The spin waves at $\Gamma$ point are collective
excitations of the same 'molecular orbital' between different blocks.

From the $2\times2$ block Hamiltonian, we obtain the eigenvalues
in the $\Gamma$ point,
\begin{eqnarray}
\begin{smallmatrix}
\omega_{1}^{(0,0)} & = & S\sqrt{(4J_{1}-J'_{1}-2J'_{2})^{2}
                        -(J'_{1}-2J'_{2})^{2}}
                        \nonumber \\
\omega_{2}^{(0,0)}=\omega_{3}^{(0,0)} & = & 2S\sqrt{(J_{1}+J_{2}
                        -J'_{2})(J_{1}-J'_{1}+J_{2}-J'_{2})}
                        \nonumber \\
\omega_{4}^{(0,0)} & = & 0\label{eq:Energy-1}
\end{smallmatrix}
\end{eqnarray}
There are four Eigenvalues. The first eigenvalue is twofold degenerate,
and its eigenvector is a combination by the first 'molecular orbital'
in the spin up and spin down block site, as shown in Fig.\ref{fig:mocular orbital}a.
The second and third eigenvalues are degenerate and each of them is
twofold degenerate. Their eigenvector are combination by the second
and third 'molecular orbital' in the spin up and spin down block site,
as shown in Fig.\ref{fig:mocular orbital}b and Fig.\ref{fig:mocular orbital}c,
respectively. They are all optical branches due to the gap
in the $\Gamma$ point. The final eigenvalue is also twofold degenerate.
However, different from the above optical branches, it is a acoustic branch
and always zero in $\Gamma$ point, which is imposed by Goldstone's
theorem.

As shown above, we can also discuss the physical properties in the
$k=(\frac{\pi}{\sqrt{10}a},\frac{\pi}{\sqrt{10}a})$ point.
From the boson Hamiltonian, we can obtain the eigenvalues in the
$k=(\frac{\pi}{\sqrt{10}a},\frac{\pi}{\sqrt{10}a})$ point,
\begin{eqnarray}
\begin{smallmatrix}
\omega_{1}^{(\pi,\pi)} & = & -2SJ_{1}+S\sqrt{(2J_{1}-J'_{1}-2J'_{2})^{2}-{J'_{1}}^{2}}
                        \nonumber \\
\omega_{2}^{(\pi,\pi)} & = & 2SJ_{1}+S\sqrt{(2J_{1}-J'_{1}-2J'_{2})^{2}-{J'_{1}}^{2}}
                        \nonumber \\
\omega_{3}^{(\pi,\pi)}=\omega_{4}^{(\pi,\pi)} & = & S\sqrt{(2J_{1}-J'_{1}
                        +2J_{2}-2J'_{2})^{2}-4{J'_{2}}^{2}-{J'_{1}}^{2}}
                        \nonumber \\
\end{smallmatrix}
\end{eqnarray}
There are also four Eigenvalues and each of them
are twofold degenerate. Six eigenvalue are optical branches and two
eigenvalue are acoustic branches. The third and forth eigenvalue are
always degenerate in $k=(\frac{\pi}{\sqrt{10}a},\frac{\pi}{\sqrt{10}a})$ point.

Now, we have five different eigenvalues for spin wave in the special
point, which can be used to fit the experimental data in order to
get the n.n. ( n.n.n.) couplings of intra- and inter-block, $J_{1}$
, $J'_{1}$ ($J_{2}$ , $J'_{2}$ ) . Then using this interaction
parameters, we can obtain the spin wave along all the BZ by numerical
diagonalization method.

\section{RESULTS AND DISCUSSION}
\label{secmodel}

In this section, we first present the phase diagram of the $J_{1}-J_{2}$
model. Then we discuss the spin wave and spin dynamical factor in
the block stipe antiferromagnetic phase.

%
\begin{figure}[tp]
\vglue -0.6cm \scalebox{1.150}[1.15]
{\epsfig{figure=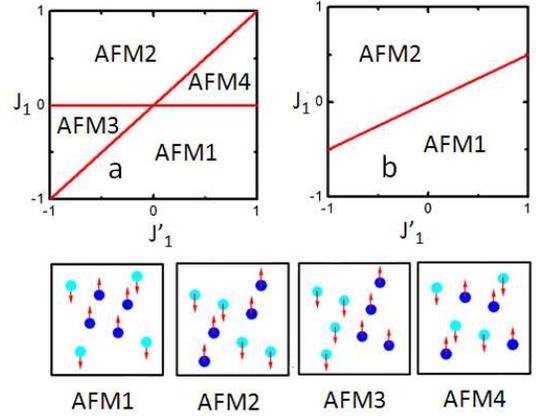,width=5.cm,angle=90}} \caption{
(color online) The phase diagram for the $J_{1}(J_{1}')-J_{2}(J_{2}')$
model. The phases are defined in Ref. 17, among which
the AFM1 phase is the block checkerboard antiferromagnetic phase observed
in the neutron diffraction experiment. The blue/azure atoms indicate
the Fe-atoms with positive/negative magnetic moment, respectively.
The magnetic configuration for AFM2, AFM3 and AFM4 is shown.
(a) The phase diagram for the interaction parameter $J_{2}=1,J_{2}'=J_{2}$.
(b) The phase diagram for the interaction parameter $J_{2}=1,J_{2}'=2.5J_{2}$.}
\label{fig:phase-diagram-1}
\end{figure}

To investigate the phase diagram for the $J_{1}-J_{2}$ Heisenberg
model, we use the stochastic Monte Carlo(MC) method to investigate the
system ground state. In the reasonable physical
parameter region, the phase diagram for the $J_{1}-J_{2}$ Heisenberg
model is given in Fig.\ref{fig:phase-diagram-1}, which is plotted
in the plane $J_{1}/J_{2}-J_{1}'/J_{2}$ at fixed value $J_{2}'/J_{2}$.
We obtain four stable phases in the case $J_{2}'/J_{2}=1$. The first
one is the block checkerboard antiferromagnetic order phase, denoted
by AFM1 in Fig.1, which has been observed by the experiment\cite{Wei-Bao-1,Wei-Bao-2}.
This phase is our interested in the study of iron-based superconductors.
Obviously, when the coupling $J_{1}$ is negative, the spin favors
to form ferromagnetic configuration in the blocks. simultaneously,
when the coupling $J_{1}'$ is positive, the spin favors to form anti-ferromagnetic
configuration between the nearest-neighbor block. Of course, when
$J_{1}'$ is negative, but small, the interaction $J_{1}$ and $J_{2}'$
are dominant and the block checkerboard antiferromagnetic phase is
also stable in this region. In the case $J_{2}'/J_{2}=1$, the block
checkerboard antiferromagnetic phase stably exists when the following
conditions are satisfied: $J_{1}<0$ and $J_{1}'>J_{1}$ . In the
parameter region $J_{1}>0$ and $J_{1}>J_{1}'$ , the system favors
to stay in the AFM2 phase, as seen in the Fig. \ref{fig:phase-diagram-1}(a).
In this phase, the antiferromagnetic order in the block arises from
the antiferromagnetic coupling $J_{1}$. On the other hand, the spin
system favors the AFM3 phase when $J_{1}'<J_{1}<0$ , which is mainly
attributed to the dominant interaction $J_{1}'$ in this parameter
regions. Similar to the phase AFM3, the system stay in the AFM4 phase
in the region $0<J_{1}<J_{1}'$, as seen in the Fig. \ref{fig:phase-diagram-1}(a).
To concretely investigate the spin properties in this system and compare
the theory calculation with experimental results, in what follows,
we change the parameter value $J_{2}'/J_{2}=1$ to $J_{2}'/J_{2}=2.5$
to investigate the phase diagram. In this parameter region, the antiferromagnetic
coupling $J_{2}'$ is dominant. Different from the first case, there
are only two phase, AFM1 and AFM2, in the phase diagram. The two phase
is separated by the line $J_{1}=0.5J_{1}'$. Below the line, the phase
is AFM1 phase, otherwise, the phase is AFM2 phase. It is interesting
to ask in which region the realistic parameters of the iron pnictides
fall. From the LDA calculations, Cao et al suggested that $J_{1}=-29$
mev, $J_{1}'=10$ mev, $J_{2}=39$ mev and $J_{2}'=95$ mev\cite{Dai-J-H}.
Such a set of parameters falls in the block checkerboard antiferromagnetic
phase in Fig.\ref{fig:phase-diagram-1}(a) and (b), implying that
the $K_{2}Fe_{4}Se_{5}$ should have the block checkerboard antiferromagnetic
order. This fact tells us we only need to focus on parameter region
in the AFM1 phase.

%
\begin{figure}[tp]
\vglue -0.6cm \scalebox{1.150}[1.15]
{\epsfig{figure=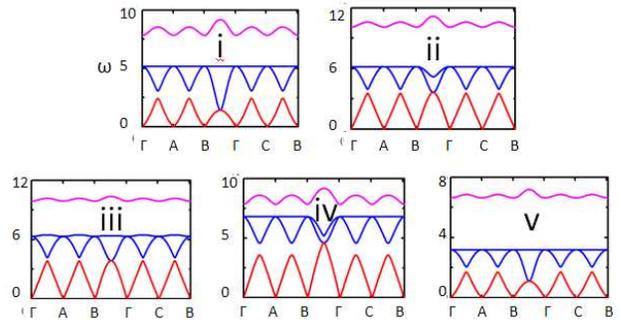,width=5.cm,angle=90}} \caption{
(color online)Acoustic and optical spin-wave branches for linear spin-wave
theory, T = 0, as a function of k along selected high symmetry directions
in the magnetic unit cell Brillouin Zone. (i)$J_{1}=-1$, $J_{1}'=0.2$, $J_{2}=1$
and $J_{2}'=2.5$, (ii)$J_{1}=-1.5$, $J_{1}'=0.2$, $J_{2}=1$ and
$J_{2}'=2.5$. , (iii) $J_{1}=-1$, $J_{1}'=1.5$, $J_{2}=1$ and
$J_{2}'=2.5$. , (iv) $J_{1}=-1$, $J_{1}'=0.2$, $J_{2}=0.2$ and
$J_{2}'=2.5$. , (v) $J_{1}=-1$, $J_{1}'=0.2$, $J_{2}=1$ and $J_{2}'=1.5$.
The x-axis correspond to the k point along the selected direction
in Fig.\ref{fig:lattice}b. y-axis correspond to the energy for spin
wave; the energy unit is $J_{2}=1$ except case (iv).}
\label{fig:The-spin-wave-line-1}
\end{figure}

First of all, we use numerical diagonalization method to investigate
the spin wave dispersion relations along the high symmetry direction
in different situations. In the numerical calculation, it is convenient
to set $J_{2}=1$; and comparing with experiments, the actual energy
scale of $J_{2}$ for the specifical material can be deduced. Motivated
by the first principle reported parameters\cite{Dai-J-H}, we first
study the set of parameter: (i)$J_{1}=-1$, $J_{1}'=0.2$, $J_{2}=1$
and $J_{2}'=2.5$. However, to study the influence of interaction
parameters on the spin-wave spectra, we also investigate four different
sets of parameter: (ii)$J_{1}=-1.5$, $J_{1}'=0.2$, $J_{2}=1$ and
$J_{2}'=2.5$. , (iii) $J_{1}=-1$, $J_{1}'=1.5$, $J_{2}=1$ and
$J_{2}'=2.5$. , (iv) $J_{1}=-1$, $J_{1}'=0.2$, $J_{2}=0.2$ and
$J_{2}'=2.5$. , (v) $J_{1}=-1$, $J_{1}'=0.2$, $J_{2}=1$ and $J_{2}'=1.5$.
In Fig. \ref{fig:The-spin-wave-line-1}, we plot spin wave dispersions
along high-symmetry direction $\Gamma-A-B-\Gamma-C-B$ in the
BZ for different interaction parameters. In all cases, there are one
acoustic and three optical spin-wave branches(each of them is twofold
degenerate). For the acoustic branches, the gap of spin wave in $\Gamma$
point is always zero due to Goldstone\textquoteright{}s theorem. In
the first case, the vibration mode for acoustic branches is shown
in Fig.\ref{fig:mocular orbital}(d). It is the collective excitation
mode for the forth 'molecular orbital' in one block with the forth
'molecular orbital' in other blocks. And the relative phase in a 'molecular
orbital' is only depend on the momentum $k$ and independent on the
interaction parameters. However, the relative phase between different
block's 'molecular orbital' is dependent on the specific interaction
parameters. The optical gap in the $\Gamma$ point is dependent on
the specific interaction parameters by contraries. As discussed in
Eq.\ref{eq:Energy-1}, two of the three optical branches are degenerate
at $\Gamma$ point. Away from $\Gamma$ point, the two degenerate
optical branches split. For example, with the increasing of k along
the $\Gamma B$ direction, the optical branch for the second
'molecular orbital' (Fig. \ref{fig:mocular orbital}b)
is almost no change. In contrast, the optical branch for the vibration
mode of third 'molecular orbital'(Fig.\ref{fig:mocular orbital}c)
has obvious dispersion, which can be clearly seen in
Fig.\ref{fig:The-spin-wave-line-1}a.
The reason that causes the splitting of spin wave can be attributed
that the vibration mode of different 'molecular orbital' is not isotropic
and dependent on the momentum $k$. Therefore, different vibration
mode has different behavior in different momentum direction. Similar
to the the above two optical branches, the third optical branch is
related to the vibration mode of the first 'molecular orbital', which
is a independent branch.

With the change of interaction parameters, we can investigate the
influence of the interaction parameters on the spin wave, as shown
in Fig.\ref{fig:The-spin-wave-line-1}. Firstly, in all cases,
the acoustic branches is always zero in $\Gamma$ point and there
are always twofold degenerate in $\Gamma$ and
$k=(\frac{\pi}{\sqrt{10}a},\frac{\pi}{\sqrt{10}a})$ point.
In the following, we compare the spin wave dispersion relation by
the change of interaction parameters. Comparing with the
Fig.\ref{fig:The-spin-wave-line-1} (i)
and Fig.\ref{fig:The-spin-wave-line-1} (ii), we find that with the increasing
of $J_{1}$, the spin wave dispersions become bigger in different
k point, especially for the acoustic branch in the
$k=(\frac{\pi}{\sqrt{10}a},\frac{\pi}{\sqrt{10}a})$ point.
From the Fig.\ref{fig:The-spin-wave-line-1} (i) and Fig.\ref{fig:The-spin-wave-line-1} (iii),
we observe that the amplitude of the second optical branch becomes
large with the increasing of $J'_{1}$. It also change the energy
for spin wave in different k point, especially in the
$k=(\frac{\pi}{\sqrt{10}a},\frac{\pi}{\sqrt{10}a})$
point. With the increasing of $J'_{1}$, the amplitude of the second
optical branch becomes more and more clear. With comparing the
Fig.\ref{fig:The-spin-wave-line-1}(i)
and Fig.\ref{fig:The-spin-wave-line-1} (iv), we observe that the interval
for the first and second optical branch becomes smaller with the increasing
of $J_{2}$. The energy for spin wave is also changed in different
k point, especially in the $k=(\frac{\pi}{\sqrt{10}a},\frac{\pi}{\sqrt{10}a})$ point.
Like all the above
cases, the spin wave dispersion also has some change in the intensity
with the increasing of $J'_{2}$. However, the interval for the first
and second optical branch becomes larger, which is one of the most
obvious features for increasing $J'_{2}$, as seen in
Fig.\ref{fig:The-spin-wave-line-1}(i) and Fig.\ref{fig:The-spin-wave-line-1} (v).

%
\begin{figure}[tp]
\vglue -0.6cm \scalebox{1.150}[1.15]
{\epsfig{figure=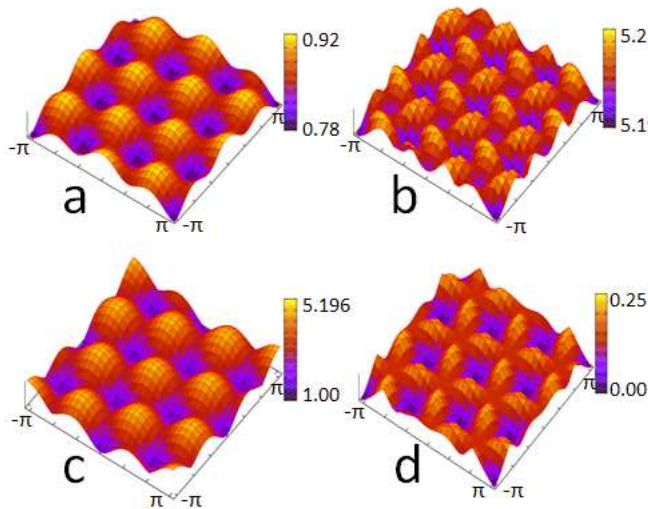,width=6.cm,angle=90}} \caption{
(color online) 3D spin wave dispersion for the parameter set of $J_{1}=-1$,
$J_{1}'=0.2$, $J_{2}=1$ and $J_{2}'=2.5$ in the extended BZ for one block, as seen
in Fig.\ref{fig:lattice} a. The energy is in units of $J_{2}$. (a)
spin wave dispersion for first optical branch. (b) spin wave dispersion
for second optical branch. (c) spin wave dispersion for third optical
branch. (d) spin wave dispersion for Acoustic branch.
The x-axis and y-axis correspond to
$3k_x-k_y$ and $k_x+3k_y$ direction respectively.}
\label{fig:The-3D-spin-wave-1}
\end{figure}

Fig.\ref{fig:The-3D-spin-wave-1} shows the typical 3 dimension spin
wave spectrum in the extended BZ for one block of the first set of parameters
in Fig.(\ref{fig:The-spin-wave-line-1}a);
and this plots provide a general qualitative overview. Regardless
of the specific parameter values, a common feature of the spin-wave
dispersion is that there are twofold degenerate in $\Gamma$
and $k=(\frac{\pi}{\sqrt{10}a},\frac{\pi}{\sqrt{10}a})$ point
and it has one zero branch in $\Gamma$ point. One acoustic branch
and three optical branch can also be seen, which is also a common
feature in this system with the fixed saddlepoint's structure. Generally
speaking, we can determine the interaction parameters by comparing
the spin-wave gap at $\Gamma$ and $k=(\frac{\pi}{\sqrt{10}a},\frac{\pi}{\sqrt{10}a})$
point with the experimental
data; then we plot the spin wave using this set of parameters, which
can be used to compare with the inelastic neutron scattering experiment.

As we all know, the neutron scattering cross section is proportional
to the dynamic structure factor $S^{in}(k,\omega)$. To further guide neutron
scattering experiment, we plot the expected neutron scattering intensity
at different constant frequency cuts in k-space. The zero-temperature
dynamic structure factor can be calculated by Holstein-Primakoff bosons.
In the linear spin-wave approximation, $S^{z}$ does not change the
number of magnons, only contributing to the elastic part of the neutron
scattering intensity. However, $S^{x}(k)$ and $S^{y}(k)$ contribute
to the inelastic neutron scattering intensity through single magnon
excitations. The spin dynamical factor associated with the spin-waves
is given by the expression,

\begin{eqnarray}
S^{in}(k,\omega) & = & S\sum_{f}|<f|\sum_{m=\pm\{1,2,3,4\}}\xi_{km}\alpha_{m}^{\dagger}|0>|^{2}
                  \nonumber \\
& & \times \delta\left(\omega-\omega_{f}\right)\label{eq:dynamical-factor}
\end{eqnarray}

Here $|0 >$ is the magnon vacuum state and $|f>$ denotes the final
state of the spin system with excitation energy. $\xi_{km}$ is the
m-th component of the eigenvector $\alpha_{m}^{\dagger}|0>$\cite{D-X-Yao-1}.

%
\begin{figure}[tp]
\vglue -0.6cm \scalebox{1.150}[1.15]
{\epsfig{figure=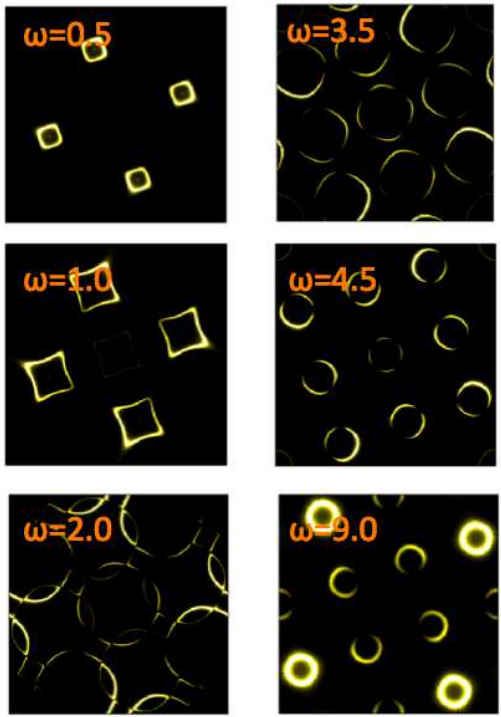,width=6.cm,angle=0}} \caption{
(Color online) Constant-energy cuts (untwinned) of the dynamic structure
factor $S^{in}(k,\omega)$ for parameters:
$J_{1}=-1$, $J_{1}'=0.2$, $J_{2}=1$ and $J_{2}'=2.5$, (ii)$J_{1}=-1.5$.
The x-axis and y-axis correspond to
$3k_x-k_y$ and $k_x+3k_y$ direction in range ($-\pi$,$\pi$) respectively.
The constant-energy
cuts from top to down = 0.5$J_{2}$ , 1.0$J_{2}$, 2.0$J_{2}$,
3.5$J_{2}$, 5.0$J_{2}$ and 9.0$J_{2}$ for the all the energy
range.}
\label{fig:The-spin-dynamical-1}
\end{figure}

Fig. \ref{fig:The-spin-dynamical-1} shows our predictions for the
intensity of the dynamical structure factor of the block checkerboard
antiferromagnetic order(an untwinned case) as a function of frequency.
At low energies($\omega=0.5$), four strongest quadrate diffraction
peaks are visible, which come from the acoustic spin-wave branch.
With the increasing of the cut frequency($\omega=1$), four strongest
quadrate diffraction peaks disperse outward toward, which is also
in the acoustic spin-wave branch range.
At intermediate frequency($\omega=2$),
the dynamic structure factor becomes a chain ring shape diffraction peaks,
which is a common results of acoustic and the third optical spin-wave branch.
However, the intensity of diffraction peaks around the $\Gamma$ point is very weak.
At high frequency($\omega=3.5$), the chain ring shape evolves to nine
strong circle diffraction peaks, which come from the two optical
spin-wave branch, which is degenerate in the $\Gamma$ point.
But the intensity of diffraction peaks around the $\Gamma$ point is also very weak.
Different from the low energy situation, in the higher frequency($\omega=4.5$),
the nine strong circle diffraction peaks don't disperse outward
toward, but inward toward; simultaneously, the middle circle diffraction
peak around the $\Gamma$ point becomes weaker and vanishes with the increasing of the cut energy.
The diffraction peak at this energy cut come from
the two optical spin-wave branches, which is degenerate in $\Gamma$
the point. At highest frequency($\omega=9.0$), four strongest circle
diffraction peaks are visible at the four corners of extended BZ for one block
and four weaker circle diffraction peaks are visible around the $\Gamma$ point and
the middle of the four boundaries,
which come from the first optical spin-wave branch. The results are
similar for other sets of parameters that have same ground states,
with the main difference being that the energy cuts must be changed
to obtain the similar spin dynamical factor patten.

\section{Discussions and Conclusions}

In conclusion, starting with the $J_{1}-J_{2}$ Heisenberg Hamiltonian
model, we have obtained the magnetic ground state phase diagram by
MQ approach and found that the block checkerboard antiferromagnetic
order is stable at reasonable physical parameter region.
In this paper, we have used
spin wave theory to investigate the spin wave and dynamic structure
factor for the block checkerboard antiferromagnetic state observed
in the iron-based superconductors. There are two acoustic branches
and six optical branches spin wave in the block checkerboard antiferromagnetic
spin system, which are the combination by the 'molecular orbital'
in the $\Gamma$ point. Then, we discussed the saddlepoint structure
in the magnetic excitation spectrum, which can also be measured by
neutron scattering experiments. The explicit analytical expressions
for the spin-wave dispersion spectra at $\Gamma$ and
$k=(\frac{\pi}{\sqrt{10}a},\frac{\pi}{\sqrt{10}a})$
have been given. Comparison with future inelastic neutron scattering
studies, we can obtain the specific values of interaction parameters.
We have also calculated the predicted inelastic neutron scattering
pattern based on linear spin wave theory. In addition, we have also
studied the specific influence of each interaction parameter on the
spin wave and dynamic structure factor. Neutron scattering experiments
about the spin wave and the behavior of spin wave at the proximity
of a quantum critical point deserve further attention.

\appendix

\section{Holstein-Primakoff Transformation \label{hp}}

The Holstein-Primakoff transformation,

\begin{eqnarray}
\hat{S}_{i}^{\dagger} & = & (\sqrt{2S-a_{i}^{\dagger}a_{i}})a_{i}\approx\sqrt{2S}a_{i}
                        \nonumber \\
\hat{S}_{i}^{-} & = & a_{i}^{\dagger}(\sqrt{2S-a_{i}^{\dagger}a_{i}})\approx\sqrt{2S}a_{i}^{\dagger}
                        \nonumber \\
\hat{S}_{i}^{z} & = & \left(S-a_{i}^{\dagger}a_{i}\right)
\end{eqnarray}

for $i=1,2,3,4$, which is occupied by spin up; S represents the size of the magnetic moment for each iron.

and

\begin{eqnarray}
\hat{S}_{-i}^{\dagger} & = & b_{-1}^{\dagger}(\sqrt{2S-b_{-i}^{\dagger}b_{-i}})\approx\sqrt{2S}b_{-i}^{\dagger}
                        \nonumber \\
\hat{S}_{-i}^{-} & = & (\sqrt{2S-b_{-i}^{\dagger}b_{-i}})b_{-i}\approx\sqrt{2S}b_{-i}
                        \nonumber \\
\hat{S}_{-i}^{z} & = & \left(b_{-i}^{\dagger}b_{-i}-S\right)
\end{eqnarray}

for $i=1,2,3,4$, which is occupied by spin down.

The fourier transform for bosonic operators is,

\begin{eqnarray}
a_{L} & = & N^{\frac{1}{2}}\Sigma_{k}e^{ik\cdot R_{L}}a_{k}
                        \nonumber \\
a_{L}^{\dagger} & = & N^{\frac{1}{2}}\Sigma_{k}e^{-ik\cdot R_{L}}a_{k}^{\dagger}
                        \nonumber \\
b_{L} & = & N^{\frac{1}{2}}\Sigma_{k}e^{ik\cdot R_{L}}b_{k}
                        \nonumber \\
b_{L}^{\dagger} & = & N^{\frac{1}{2}}\Sigma_{k}e^{-ik\cdot R_{L}}b_{k}^{\dagger}
\end{eqnarray}

Here, we define N is the number of the magnetic unit cell. $L$ is
the site index and k is momentum index.

\section{Spin Wave Hamiltonian \label{spin-wave}}

We write the detailed expression in Eq. 2. Replacing the
spin operators by HP bosons, we can get the HP boson Hamiltonian $\mathcal{H}_{k}$,

\begin{eqnarray*}
\begin{smallmatrix}
\left\{ \begin{array}{cccccccc}
E_{1} & F_{k} & G_{k} & S_{k} & 0 & \left(B_{k}\right)^{*} & \left(C_{k}\right)^{*} & \left(A_{k}\right)^{*}\\
\left(F_{k}\right)^{*} & E_{1} & S_{k} & T_{k} & B_{k} & 0 & \left(A_{k}\right)^{*} & \left(D_{k}\right)^{*}\\
\left(G_{k}\right)^{*} & \left(S_{k}\right)^{*} & E_{1} & (F_{k})^{*} & C_{k} & A_{k} & 0 & B_{k}\\
\left(S_{k}\right)^{*} & \left(T_{k}\right)^{*} & F_{k} & E_{1} & A_{k} & D_{k} & \left(B_{k}\right)^{*} & 0\\
0 & (B_{k})^{*} & (C_{k})^{*} & (A_{k})^{*} & E_{1} & F_{k} & G_{k} & S_{k}\\
B_{k} & 0 & (A_{k})^{*} & (D_{k})^{*} & (F_{k})^{*} & E_{1} & S_{k} & T_{k}\\
C_{k} & A_{k} & 0 & B_{k} & (G_{k})^{*} & (S_{k})^{*} & E_{1} & \left(F_{k}\right)^{*}\\
A_{k} & D_{k} & (B_{k})^{*} & 0 & (S_{k})^{*} & (T_{k})^{*} & F_{k} & E_{1}\end{array}\right\}
\label{eq:A1}
\end{smallmatrix}
\end{eqnarray*}
and
\begin{eqnarray*}
\begin{smallmatrix}
E_{k} & = & J_{1}+1J'_{1}-1J_{2}+2J'_{2}
                        \nonumber \\
A_{k} & = & J'_{2}Se^{-i\overrightarrow{k}\cdot
            \overrightarrow{x_{a}}}
                        \nonumber \\
B_{k} & = & J'_{2}Se^{i\overrightarrow{k}\cdot \overrightarrow{y_{a}}}
                        \nonumber \\
C_{k} & = & J'_{1}Se^{-i\overrightarrow{k}\cdot(0.5\overrightarrow{x_{a}}+0.5\overrightarrow{y_{a}})}
                        \nonumber \\
D_{k} & = & J'_{1}Se^{i\overrightarrow{k}\cdot(0.5\overrightarrow{x_{a}}-0.5\overrightarrow{y_{a}})}
                        \nonumber \\
F_{k} & = & SJ_{1}e^{-i\overrightarrow{k}\cdot(0.5\overrightarrow{x_{a}}-0.5\overrightarrow{y_{a}})}
                        \nonumber \\
G_{k} & = & SJ_{2}e^{-i\overrightarrow{k}\cdot \overrightarrow{x_{a}}}
                        \nonumber \\
S_{k} & = & SJ_{1}e^{-i\overrightarrow{k}\cdot(0.5\overrightarrow{x_{a}}+0.5\overrightarrow{y_{a}})}
                        \nonumber \\
T_{k} & = & SJ_{2}e^{-i\overrightarrow{k}\cdot \overrightarrow{y_{a}}}
                        \nonumber \\
\overrightarrow{x_{a}} & = & \sqrt{1.6}a\overrightarrow{i}+\sqrt{0.4}a\overrightarrow{j}
                        \nonumber \\
\overrightarrow{y_{a}} & = & -\sqrt{0.4}a\overrightarrow{i}+\sqrt{1.6}a\overrightarrow{j}
\end{smallmatrix}
\end{eqnarray*}
In the 'molecular orbital' basis, the Hamiltonian $\mathcal{H}^{orbital}_{k}$ becomes,
\begin{eqnarray*}
\begin{smallmatrix}
\frac{1}{4}\left\{ \begin{array}{cccccccc}
E_{o}^{1} & 0 & 0 & 0 & A_{o} & B_{o} & C_{o} & D_{o}\\
0 & E_{o}^{2} & 0 & 0 & \left(B_{o}\right)^{*} & F_{o} & -D_{o} & G_{o}\\
0 & 0 & E_{o}^{3} & 0 & \left(C_{o}\right)^{*} & -D_{o} & P_{o} & S_{o}\\
0 & 0 & 0 & E_{o}^{4} & D_{o} & \left(G_{o}\right)^{*} & \left(S_{o}\right)^{*}
& T_{o}\\
A_{o}^{4} & B_{o} & C_{o} & D_{o} & E_{o}^{1} & 0 & 0 & 0\\
\left(B_{o}\right)^{*} & F_{o} & -D_{o} & G_{o} & 0 & E_{o}^{2} & 0 & 0\\
\left(C_{o}\right)^{*} & -D_{o} & P_{o} & S_{o} & 0 & 0 & E_{o}^{3} & 0\\
D_{o} & \left(G_{o}\right)^{*} & \left(S_{o}\right)^{*}
& T_{o} & 0 & 0 & 0 & E_{o}^{4}\end{array}\right\}
\end{smallmatrix}
\end{eqnarray*}
and
\begin{eqnarray*}
\begin{smallmatrix}
E_{o}^{1} & = & -16J_{1}+4J'_{1}+8.J'_{2}
                        \nonumber \\
E_{o}^{2} & = & -8J_{1}+4J'_{1}-8J_{2}+8J'_{2}
                        \nonumber \\
E_{o}^{3} & = & -8J_{1}+4J'_{1}-8J_{2}+8J'_{2}
                        \nonumber \\
E_{o}^{4} & = & 4J'_{1}+8J'_{2}\\
A_{o} & = & [2J'_{1}-4J'_{2}].[cos(a_{o})+cos(b_{o})]
                        \nonumber \\
B_{o} & = & -2J'_{1}[sin(a_{o})+sin(b_{o})]i
                        \nonumber \\
 &  & +4J'_{2}sin(b_{o})i
                        \nonumber \\
C_{o} & = & -2J'_{1}[sin(a_{o})-sin(b_{o})]i
                        \nonumber \\
 &  & +4J'_{2}sin(a_{o})i
                        \nonumber \\
D_{o} & = & 2J'_{1}[cos(a_{o})-cos(b_{o})]
                        \nonumber \\
F_{o} & = & -2J'_{1}[cos(a_{o})+cos(b_{o})]
                        \nonumber \\
 &  & -4J'_{2}[cos(a_{o})-cos(b_{o})]
                        \nonumber \\
G_{o} & = & 2J'_{1}[sin(a_{o})-sin(b_{o})]i
                        \nonumber \\
 &  & +4J'_{2}sin(a_{o})i
                        \nonumber \\
P_{o} & = & 4J'_{2}[cos(a_{o})-cos(b_{o})]]
                        \nonumber \\
 &  & -2J'_{1}[cos(a_{o})+cos(b_{o})
                        \nonumber \\
S_{o} & = & 4J'_{2}sin( b_{o})i
                        \nonumber \\
 &  & +2J'_{1}[sin(a_{o})+sin( b_{o})]i
                        \nonumber \\
T_{o} & = & [2J'_{1}+4J'_{2}].[cos(a_{o})+cos( b_{o})]
                         \nonumber \\
 a_{o} & = & 1.5\overrightarrow{k}\cdot \overrightarrow{x_{a}}+0.5\overrightarrow{k}\cdot \overrightarrow{y_{a}}
                          \nonumber \\
 b_{o} & = & 0.5\overrightarrow{k}\cdot \overrightarrow{x_{a}}-1.5\overrightarrow{k}\cdot \overrightarrow{y_{a}}
\end{smallmatrix}
\end{eqnarray*}

The matrix elements for different 'molecular orbital' in the same
block is zero.
\section{Numerical Diagonalization Method \label{num-dig}}
We can use the numerical method to diagonalize the boson
pairing Hamiltonian. The matrix form for boson Hamiltonian is,

\begin{eqnarray*}
\psi^{\dagger}\hat{\mathbb{\mathcal{H}}}\psi & = & \left[a^{\dagger},b\right]\left[\begin{array}{cc}
A & B\\
B & A\end{array}\right]\left[\begin{array}{c}
a\\
b^{\dagger}\end{array}\right]\end{eqnarray*}

Here, $\psi^{\dagger}=\left(a^{\dagger},b\right)=\left(a_{1}^{\dagger},a_{2}^{\dagger},a_{3}^{\dagger},a_{4}^{\dagger},b_{-1},b_{-2},b_{-3},b_{-4}\right)$,
A and B is a $4\times4$ matrix; a and b are boson operators. The operators satisfy the commutation relation,

\begin{eqnarray*}
\left[\psi_{i},\psi_{j}^{\dagger}\right] & = & \hat{I}_{-i,j}
\end{eqnarray*}

and

\begin{eqnarray*}
\hat{I}_{-} & = & \left(\begin{array}{cc}
I & 0\\
0 & -I\end{array}\right)\end{eqnarray*}

$I$ is a $4\times4$ identity matrix.

The diagonalization problem amounts to finding a transformation $T$,
which lets $\hat{T}^{\dagger}\hat{\mathcal{H}}\hat{T}$ become a diagonalization
matrix $\Omega$.

\begin{eqnarray*}
\psi & = & \left[\begin{array}{c}
a\\
b^{\dagger}\end{array}\right]=\hat{T}\left[\begin{array}{c}
\alpha\\
\beta^{\dagger}\end{array}\right]=\hat{T}\varphi\end{eqnarray*}

Here, we require that $\alpha$ and $\beta$ are also a set of boson
operators and $\varphi^{\dagger}=\left(\alpha^{\dagger},\beta\right)=\left(\alpha_{1}^{\dagger},\alpha_{2}^{\dagger},\alpha_{3}^{\dagger},\alpha_{4}^{\dagger},\beta_{-1},\beta_{-2},\beta_{-3},\beta_{-4}\right)$.

Then,

\begin{eqnarray*}
\left[\psi_{i},\psi_{j}^{\dagger}\right] & = & \hat{I}_{-i,j}=\sum_{i',j'}T_{i,i'}I_{-i',j'}T_{j',j}^{\dagger}\end{eqnarray*}

we get

\begin{equation}
\hat{T}\hat{I}_{-}\hat{T}^{\dagger}=\hat{I}_{-}\label{eq:Dig-2}\end{equation}

Since $\hat{I}_{-}^{2}=I$. we know $I=(\hat{I}_{-}\hat{T})(\hat{I}_{-}\hat{T}^{\dagger})$.
Due to $(\hat{I}_{-}\hat{T})$ and $(\hat{I}_{-}\hat{T}^{\dagger})$
are each inverses, thus $I=(\hat{I}_{-}\hat{T}^{\dagger})(\hat{I}_{-}\hat{T})$
and $\hat{T}^{\dagger}\hat{I}_{-}\hat{T}=\hat{I}_{-}$.

Then we want the final form of $\hat{\mathcal{H}}$ is diagonalization.

\begin{eqnarray}
\hat{\mathcal{H}} & = & \left[a^{\dagger},b\right]\left[\begin{array}{cc}
A & B\\
B & A\end{array}\right]\left[\begin{array}{c}
a\\
b^{\dagger}\end{array}\right]\nonumber \\
 & = & \psi^{\dagger}\hat{\mathbb{\mathcal{H}}}\psi\nonumber \\
 & = & \varphi^{\dagger}\left\{ \hat{T}^{\dagger}\left[\begin{array}{cc}
A & B\\
B & A\end{array}\right]\hat{T}\right\} \varphi\nonumber \\
 & = & \left[\alpha^{\dagger},\beta\right]\left\{ \hat{T}^{\dagger}\left[\begin{array}{cc}
A & B\\
B & A\end{array}\right]\hat{T}\right\} \left[\begin{array}{c}
\alpha\\
\beta^{\dagger}\end{array}\right]\nonumber \\
 & = & \left[\alpha^{\dagger},\beta\right]\left\{ \hat{\Omega}\right\} \left[\begin{array}{c}
\alpha\\
\beta^{\dagger}\end{array}\right]\nonumber \\
 & = & \varphi^{\dagger}\hat{\Omega}\varphi\nonumber \\
 & = & \sum_{i}^{4}\omega_{i}\alpha^{\dagger}\alpha+\omega_{-i}\beta\beta^{\dagger}\label{eq:Dig-1}\end{eqnarray}

Here,$\hat{\Omega}=diag\left(\omega_{1},\cdots,\omega_{4},\omega_{-1},\cdots,\omega_{-4}\right)$
represents a diagonalization matrix and the matrix elements for diagonalization
matrix is $\left(\omega_{1},\cdots,\omega_{4},\omega_{-1},\cdots,\omega_{-4}\right)$
.

In other words, we want the expression $\hat{T}^{\dagger}\mathcal{\hat{H}}\hat{T}=\Omega$
and matrix $\hat{\Omega}$ is a diagonalization matrix. We must solve
the matrix $\hat{T.}$

Combining Eq.\ref{eq:Dig-2} and Eq.\ref{eq:Dig-1}, we get,

\begin{eqnarray}
\hat{T}^{\dagger}\mathcal{\hat{H}}\hat{T} & = & \hat{\Omega}\nonumber \\
\left[\hat{T}\hat{I}_{-}\right]\hat{T}^{\dagger}\mathcal{\hat{H}}\hat{T} & = & \left[\hat{T}\hat{I}_{-}\right]\hat{\Omega}\nonumber \\
\left[\hat{T}\hat{I}_{-}\hat{T}^{\dagger}\right]\mathcal{\hat{H}}\hat{T} & = & \left[\hat{T}\hat{I}_{-}\right]\hat{\Omega}\nonumber \\
\hat{I}_{-}\mathcal{\hat{H}}\hat{T} & = & \left[\hat{T}\hat{I}_{-}\right]\hat{\Omega}\nonumber \\
(\hat{I}_{-}\hat{\mathcal{H}})\hat{T} & = & \hat{T}\left[\hat{I}_{-}\hat{\Omega}\right]\nonumber \\
(\hat{I}_{-}\hat{\mathcal{H}})\hat{T} & = & \hat{T}\left[\hat{\lambda}\right]\label{eq:Dig-3}\end{eqnarray}

Here $\hat{\lambda}=\left[\hat{I}_{-}\hat{\Omega}\right]=diag\left(\omega_{1},\cdots,\omega_{4},-\omega_{-1},\cdots,-\omega_{-4}\right)$
is a diagonalization matrix. In other words, if we want to get $\hat{T}^{\dagger}\mathcal{\hat{H}}\hat{T}=\hat{\Omega}$,
we can solve the general Hamiltonian $(\hat{I}_{-}\hat{\mathcal{H}})\hat{T}=\hat{T}\left[\hat{\lambda}\right]$.

J.L. van Hemmen's\cite{M-W-Xiao-1} strategy is that the canonical
transformation $\hat{T}$ is fully determined by its n(=8) columns
\{$x_{1}$, ... , $x_{1}$, $x_{-1}$, ... , $x_{-4}$\}. We, therefore,
reduce Eq.\ref{eq:Dig-3} to an eigenvalue problem for these n(=8)
vectors $x_{i}$, $1\leqq i\leqq8$.

Then

\begin{eqnarray*}
\left(\hat{I}_{-}\hat{\mathcal{H}}\right)\chi & = & \lambda\chi\end{eqnarray*}

Where $\chi=x_{i}$, $1\leqq i\leqq8$, and $\lambda\in\left\{ \omega_{1},\omega_{2},\omega_{3},\omega_{4},-\omega_{-1},-\omega_{-2},-\omega_{-3},-\omega_{-4}\right\} $.
So, to diagonalize the Hamiltonian is equivalence to solve a general
eigenvalue problem.

If we define a matrix $\hat{I}_{y}=\left[\begin{array}{cc}
0 & I\\
I & 0\end{array}\right]$ and $I$ is a $4\times4$ identity matrix, it is easy to proof ,

\begin{eqnarray*}
\hat{I}_{-}\mathcal{\hat{H}} & = & -\hat{I}_{y}^{-1}\hat{I}_{-}\mathcal{\hat{H}}\hat{I}_{y}\end{eqnarray*}

We assume there is a eigenvalue $\lambda_{i}$ and eigenvector $\chi_{i}=\left[\begin{array}{c}
\mu_{1-4}\\
\nu_{1-4}\end{array}\right]$ for the general Hamiltonian,

\begin{eqnarray*}
\hat{I}_{-}\hat{\mathcal{H}}\chi_{i} & = & \lambda_{i}\chi_{i}\end{eqnarray*}

Then the $\chi'_{i}=\hat{I}_{y}\cdot\chi_{i}=\left[\begin{array}{c}
\nu_{1-4}\\
\mu_{1-4}\end{array}\right]$ is also eigenvector of $\hat{I}_{-}\hat{\mathcal{H}}$ and the corresponding
eigenvalue is $\lambda'=-\lambda$:

\begin{eqnarray*}
\hat{I}_{-}\hat{\mathcal{H}}\chi_{i} & = & \lambda_{i}\chi_{i}\\
-\hat{I}_{y}^{-1}\hat{I}_{-}\hat{\mathcal{H}}\left[\hat{I}_{y}\chi_{i}\right] & = & \lambda_{i}\chi_{i}\\
\hat{I}_{y}\hat{I}_{y}^{-1}\hat{I}_{-}\hat{\mathcal{H}}\left[\hat{I}_{y}\chi_{i}\right] & = & -\hat{I}_{y}\lambda_{i}\chi_{i}\\
\hat{I}_{-}\hat{\mathcal{H}}\left[\hat{I}_{y}\chi_{i}\right] & = & -\lambda_{i}\hat{I}_{y}\chi_{i}\\
I_{-}\hat{\mathcal{H}}\chi'_{i} & = & -\lambda_{i}\chi'_{i}\\
I_{-}\hat{\mathcal{H}}\chi'_{i} & = & \lambda'_{i}\chi'_{i}
\end{eqnarray*}
Therefore, the eigenvalue are in pairs. For the sake of convenience,
we arrange the order of eigenvalues by the relative size of the value
$\aleph_{i}=\mid\mu_{i}\mid^{2}-\mid\nu_{i}\mid^{2}$, ($\aleph_{1}>......>\aleph_{8}$);
for the same $\aleph_{i}$, we arrange the order of eigenvalues by
its relative size. If the eigenvalue $\lambda_{i}$ for the corresponding
eigenvector $\aleph_{1}>\aleph_{2}>\aleph_{3}>\aleph_{4}$ is lower
than zero, it represents the ground state is not stable. For the first
four eigenvector $\mid\mu_{i}\mid^{2}-\mid\nu_{i}\mid^{2}=1$(i=1,2,3,4)
and for the last four eigenvector $\mid\mu_{i}\mid^{2}-\mid\nu_{i}\mid^{2}=-1$
(i=5,6,7,8).


\begin{thebibliography}{}
\bibitem{Lee-P-A-1}P. A. Lee, N. Nagaosa, and X. G. Wen, Rev. Mod.
Phys. 78, 17 (2006).

\bibitem{Bednorz-1}J. G. Bednorz, and K. A. Muller, Z. Phys. B 64,
189 (1986).

\bibitem{Kamihara-1} Y. Kamihara, T. Watanabe, M. Hirano, and H.
Hosono, J. Am. Chem. Soc. 130 3296 (2008).

\bibitem{Chen-G-F-1}G. F. Chen, Phys. Rev. Lett. 101 057007 (2008).

\bibitem{Lei-H-1}Hechang Lei, KefengWang, J. B. Warren, andC. Petrovic,
arXiv:1102.2215.

\bibitem{Kondo-1}Takeshi Kondo, A. F. Santander-Syro, O. Copie, Chang
Liu, M. E. Tillman, E. D. Mun, J. Schmalian, S. L. Bud\textquoteright{}ko,
M. A. Tanatar, P. C. Canfield, and A. Kaminski, Phys. Rev. Lett.,
101, 147003 (2008).

\bibitem{Wray-1}L. Wray, D. Qian, D. Hsieh, Y. Xia, L. Li, J. G.
Checkelsky, A. Pasupathy, K. K. Gomes, C. V. Parker, A. V. Fedorov,
G. F. Chen, J. L. Luo, A. Yazdani, N. P. Ong, N. L. Wang, and M. Z.
Hasan, Phys. Rev. B, 78, 184508 (2008).

\bibitem{Xu-Gang-1}Gang Xu, Haijun Zhang, Xi Dai, and Zhong Fang,
EPL, 84, 67015 (2008).

\bibitem{Borisenko-1}S. V. Borisenko, V. B. Zabolotnyy, D. V. Evtushinsky,
T. K. Kim, I. V. Morozov, A. N. Yaresko, A. A. Kordyuk, G. Behr, A.
Vasiliev, R. Follath, and B. Bhner, Phys. Rev. Lett. 105, 067002
(2010) .

\bibitem{Khasanov-1}R. Khasanov, M. Bendele, A. Amato, K. Conder,
H. Keller, H.-H. Klauss, H. Luetkens, and E. Pomjakushina, Phys. Rev.
Lett. 104, 087004 (2010) .

\bibitem{Guo-2}J. Guo et al., arXiv:1101.0092 (2011); Y. Kawasaki
et al., arXiv:1101.0896 (2011); J. J. Ying et al., arXiv:1101.1234
(2011).

\bibitem{Maziopa-1}A. Krzton-Maziopa et al., arXiv:1012.3637 (2010).
J. J. Ying et al., arXiv:1012.5552 (2010). A. F. Wang et al., arXiv:1012.5525
(2010) M. Fang et al., arXiv:1012.5236 (2010); H. Wang et al., arXiv:1101.0462
(2011).

\bibitem{Y-Z-Lu-1}Xun-WangYan, MiaoGao, Zhong-Yi Lu, and TaoXiang,
arXiv:1102.2215.

\bibitem{Wei-Bao-3}Wei Bao, G. N. Li, Q. Huang, G. F. Chen, J. B.
He, M. A. Green, Y. Qiu, D. M. Wang, J. L. Luo, andM. M. Wu, arXiv:1102.3674.

\bibitem{Wei-Bao-1} Wei Bao, Q. Huang, G. F. Chen, M. A. Green, D.
M. Wang, J. B. He X. Q. Wang andY. Qiu, arXiv:1102.0830.

\bibitem{Wei-Bao-2}F. Ye, S. Chi, Wei Bao, X. F. Wang, J. J. Ying,
X. H. Chen, H. D. Wang, C. H. Dong, and Minghu Fang,arXiv:1102.2282.

\bibitem{Dai-J-H} C. Cao and J. Dai (2011), arXiv:1102.1344.

\bibitem{Yu-V-1}V. Yu. Pomjakushin, E. V. Pomjakushina, A. Krzton-Maziopa,
K. Conder, and Z. Shermadini, arXiv:1102.3380.

\bibitem{You-Y-Z-1}Yi-ZhuangYou, FangYang, Su-PengKou, and Zheng-YuWeng,
arXiv:1102.3200.

\bibitem{Zhang-G-M-1}G. M. Zhang, Z. Y. Lu, and T. Xiang, arXiv:1102.4575
(2011).

\bibitem{Auerbach-1}A. Auerbach, Phys. Rev. Lett., 72, 2931 (1994).

\bibitem{D-X-Yao-1}E. W. Carlson, D. X. Yao, and D. K. Campbell,
Phys. Rev. B , 70, 064505 (2004); D. X. Yao, E. W. Carlson, and D.
K. Campbell, Phys. Rev. Lett. 97, 017003 (2006).

\bibitem{M-W-Xiao}Ming-wen Xiao, arXiv:0908.0787.

\bibitem{M-W-Xiao-1}J.L. van Hemmen, Z. Physik B-Condensed Matter
38, 271-277 (1980).

\bibitem{M-W-Xiao-3}Lu Huaixin and Zhang Yongde, International Journal
of Theoretical Physics, 39, 2, (2000).

\bibitem{M-W-Xiao-4}Constantino Tsallis, J. Math. Phys. 19, 277,
(1978).

\bibitem{M-W-Xiao-5}J.H.P. ColpaCOLPA, Physica A, 93, 327, (1978).
\end{thebibliography}
\end{document}